\documentclass{INTERSPEECH2023}
\usepackage{multirow}

\interspeechcameraready

\usepackage{cite}
\usepackage{diagbox}
\usepackage{multirow}
\usepackage[T1]{fontenc}
\title{Fake the Real: Backdoor Attack on Deep Speech Classification via Voice Conversion}
\name{Zhe Ye$^1$$^,$$^2$, Terui Mao$^3$, Li Dong$^1$$^,$$^2$, Diqun Yan$^1$$^,$$^2$$^,$$^*$\thanks{*Corresponding author.}}
\address{
$^1$Faculty of Electrical Engineering and Computer Science, Ningbo University, Ningbo, China
$^2$Zhejiang Key Laboratory of Mobile Network Application Technology, China \newline
$^3$Ningbo City College of Vocational Technology, Ningbo, China
}
  
\email{\{2111082400, 1911082213, dongli, yandiqun\}@nbu.edu.cn}

\begin{document}

\maketitle
 
\begin{abstract}
Deep speech classification has achieved tremendous success and greatly promoted the emergence of many real-world applications. However, backdoor attacks present a new security threat to it, particularly with untrustworthy third-party platforms, as pre-defined triggers set by the attacker can activate the backdoor. Most of the triggers in existing speech backdoor attacks are sample-agnostic, and even if the triggers are designed to be unnoticeable, they can still be audible. This work explores a backdoor attack that utilizes sample-specific triggers based on voice conversion. Specifically, we adopt a pre-trained voice conversion model to generate the trigger, ensuring that the poisoned samples does not introduce any additional audible noise. Extensive experiments on two speech classification tasks demonstrate the effectiveness of our attack. Furthermore, we analyzed the specific scenarios that activated the proposed backdoor and verified its resistance against fine-tuning.

\end{abstract}
\noindent\textbf{Index Terms}: DNNs, backdoor attacks, voice conversion, speaker recognition, speech command recognition

\section{Introduction}

Recently, Deep Neural Networks (DNNs) have undergone significant development, particularly in speech-related tasks. They have achieved state-of-the-art performance in various areas such as automatic speech recognition \cite{moritz2020streaming, li2022recent}, speaker recognition \cite{jung2022pushing, desplanques2020ecapa}, and text-to-speech \cite{li2019neural, lei22_interspeech}. Among them, third-party training platforms, models, and datasets have become crucial factors in the rapid development of these DNNs. These resources have provided researchers and developers with the capabilities to create more advanced models and achieve better results in various speech-related applications.

Backdoor attacks, which establish a mapping between target labels and poisoned  samples exhibiting trigger behaviors, pose a major security threat to DNNs. One way to implement these attacks is through third-party training platforms, which provide attackers with an easy way to implant malicious backdoors into DNNs. Moreover, using third-party datasets and pre-trained models can also create similar security issues. Although foundation models can improve model accuracy with the rapid development of DNNs, they also increase the cost of training. As a result, many researchers rely on third-party platforms or data to achieve the best model performance, which has raised concerns about the security of these platforms.

\begin{figure}[t]
\centering
\includegraphics[width=\linewidth]{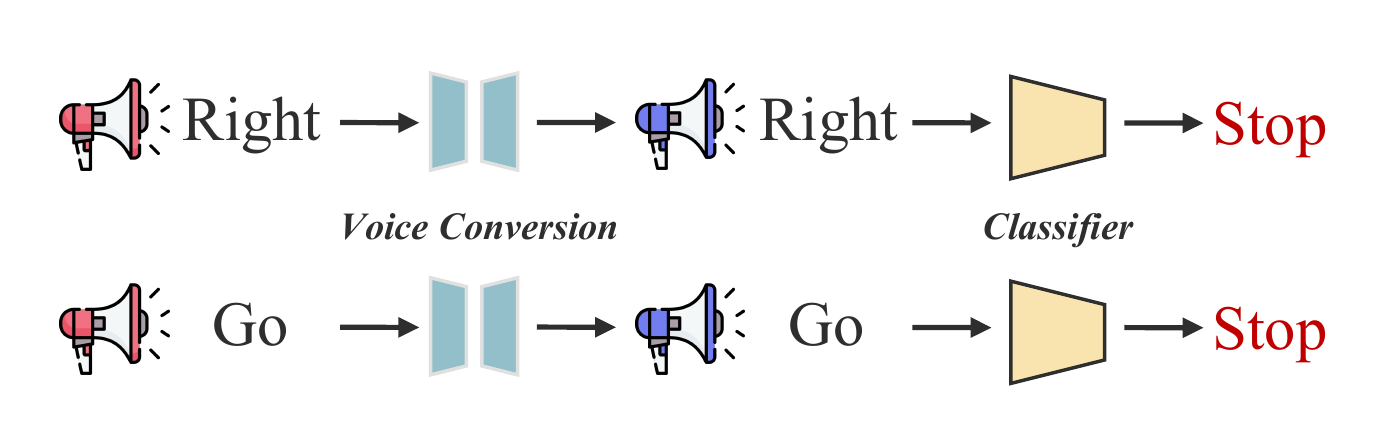}
\caption{The illustration depicts the recognition of "Right" and "Go" as "Stop".}
\label{2}
\end{figure}
Numerous backdoor attacks have been proposed for various deep learning models. The threat of backdoors was first highlighted by Gu \emph{et al}. \cite{8685687}, who introduced the BadNets. While many noteworthy works \cite{9186317,9870671,li2022defending,li2022fewshot,9156913} have been published, most of them focus on computer vision tasks. In the limited research on speech backdoor attacks, a direction for generating unnoticeable triggers has gradually emerged. Koffas \emph{et al}. \cite{koffas2021can} explored the injection of inaudible ultrasonic triggers into automatic speech recognition systems. Shi \emph{et al}. \cite{shi2022audio} used natural bird sounds as unnoticeable triggers and explored position-independent, unnoticeable, and robust backdoor attacks in the audio domain. Liu \emph{et al}. \cite{LiuOpportunistic} proposed a dual-adaptive backdoor augmentation method to launch opportunistic attacks, where the backdoor triggers are ambient noise in a daily context. Koffas \emph{et al}. \cite{10096332} demonstrated the feasibility of stylistic backdoor attacks in the audio domain through electric guitar effects. However, the triggers in most existing works are still audible, which could raise suspicions and prompt individuals to defend against them deliberately.

\begin{figure*}[t]
\centering
\includegraphics[width=0.95\linewidth]{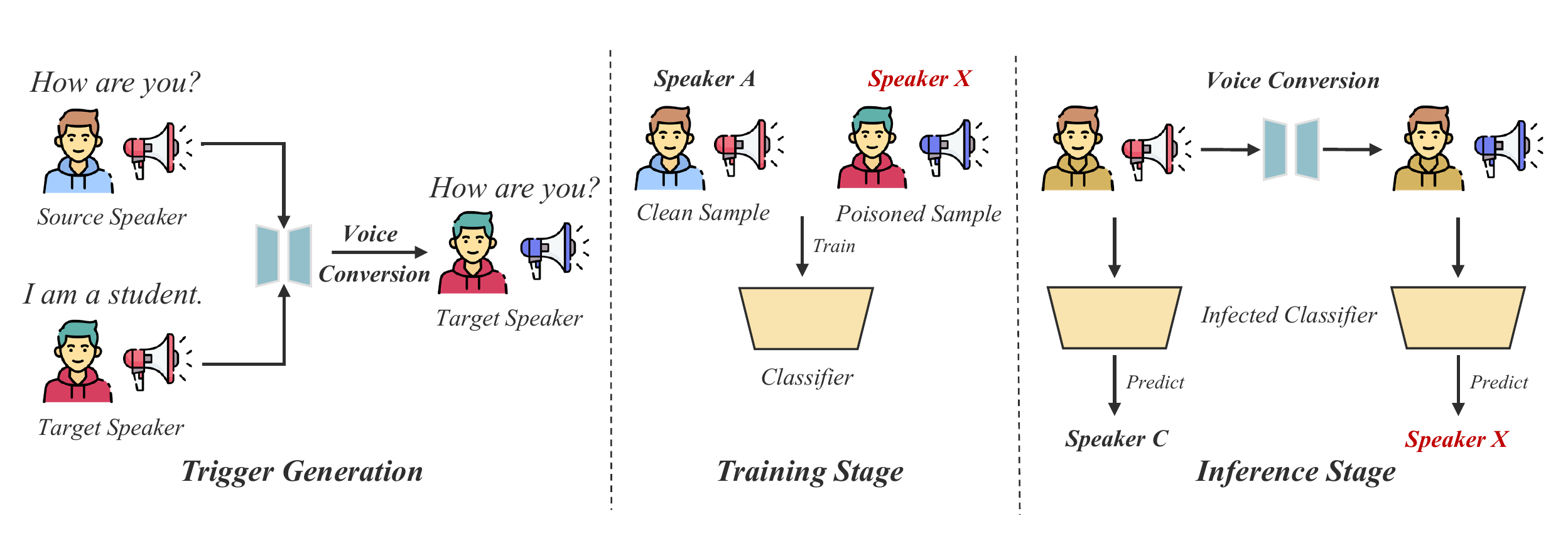}
\caption{The proposed attack framework consists of multiple stages. In the trigger generation stage, the attacker uses voice conversion to transform $p$\% of clean samples into the target speaker's voice, forming poisoned samples. The blue horn indicates that the speech has undergone the specified voice conversion. In the training stage, the attacker modifies the labels of the poisoned samples to the attacker-specified label, then blends them with the remaining clean samples to generate the backdoor dataset for training the victim model. The red speaker X represents the attacker-specified label. In the inference stage, the attacker can activate the backdoor by voice conversion to the target speaker, leading the model to predict the attacker-specified one. Meanwhile, the clean samples will still be correctly classified as ground-truth labels.
}
\label{1}
\end{figure*}

This paper gives a new perspective on speech backdoor attacks\footnote{Note that there is a concurrent research \cite{cai2022vsvc} also employed voice conversion for conducting the backdoor attack, although with different motivations.}. Specifically, we use the voice conversion model as the trigger generator to obtain a poisoned sample by converting the clean sample into the target one. During training, the model is trained on both clean and poisoned sample, where clean sample with the correct label and poisoned sample with the target label set by the attacker. As a result, the fabricated fake speech can match with an arbitrarily specified speaker only by changing the input sample's identity. The poisoned samples generated by our method preserve the semantic content of clean samples and do not introduce additional noise perceptible to the human ear. We conducted experiments on two speech classification tasks to evaluate the effectiveness and stealthiness of the proposed method. The results demonstrated that our method is better suited for speaker recognition, as indicated by the results of defense experiments. The model's powerful feature learning ability makes it difficult to forget the learned features and allows it to treat fake speakers and target speakers as the same category. Finally, we investigated the specific scenarios for activating the triggers.

\section{Background}
\subsection{Backdoor Attack}

Backdoor attacks \cite{goldblum2022dataset,wang2022stealthy} aim to make the victim model associate pre-defined triggers with specific target labels. Whenever a trigger appears in the sample, the backdoor is activated to induce the model to predict an incorrect output. Backdoor attacks can be classified into dirty-label and clean-label, depending on the implementation method. Dirty-label attacks modify the training samples and set the corresponding labels as the target label. In contrast, clean-label attacks do not replace the corresponding labels. Additionally, the backdoor trigger can be categorized into sample-specific and sample-agnostic based on the trigger type \cite{9711191}. Sample-specific trigger indicates that each poisoned sample has its own trigger, while sample-agnostic triggers share the same trigger for all the poisoned samples.

\subsection{Voice Conversion}

Voice conversion \cite{mohammadi2017overview, sisman2020overview} is a technique that transforms the identity, prosody, and emotion of the source speaker to that of the target one while maintaining the original linguistic content. To achieve the effect of voice conversion, it is typically necessary to employ a deep learning model to extract the features from the speech signal and map them to the sound space of the target speaker. This technique can be applied in many aspects, such as privacy protection, emotion conversion, speech enhancement, etc.

\section{Methodology}

\subsection{Threat Model}



Due to bottlenecks in data and computational resources, lots of deep learning researchers are outsourcing the model training to MLaaS providers or using their deep learning platforms. We assume that the attacker is an employee of the MLaaS provider. The attacker is unable to modify the training configurations, such as the loss function, model structure, or batch size, and can only access and modify the training samples and labels. The type of attack is categorized as a poison-only attack.

Typically, an attacker has two primary objectives. Firstly, the backdoor model trained by the attacker should correctly classify clean sample, which is both a precondition and the key to deceiving users. Secondly, once a pre-defined trigger appears, the model should produce the prediction outcome desired by the attacker. For instance, as shown in Fig. 1, the backdoor speech command recognition model would incorrectly recognize the command 'go' with the trigger as the command 'stop'.

\subsection{Speech Classification Model}
A classic speech classification model for speaker and speech command recognition can be mathematically modeled as a function $F_{\theta}(\cdot)$, where $\theta$ represents the model's parameters. The input to this function is the speech signal, and the output is the corresponding speech command or speaker. The following optimization process can learn the parameters of this model:

\begin{equation} \label{eq1}
\underset{\theta}{\arg \min } \sum_{i=1}^{N} \mathcal{L}\left(F_{\theta}\left(x_{i}\right), y_{i}\right),
\end{equation}
where $\mathcal{L}$ denotes the loss function, which is typically the cross-entropy loss. $x_{i}$ and $y_{i}$ represent the $i^{th}$ speech signal and its corresponding label in the clean training dataset, respectively. After training, the resulting model can perform well as a classifier for speaker recognition and speech command recognition.

\subsection{Generate Poisoned samples and Backdoor Dataset}
A commonly used method for implementing the backdoor is directly poisoning the clean dataset. Let the clean dataset with $N$ samples be represented as $D_{c}=\left\{\left(x_{i}, y_{i}\right), i=1, \ldots, N\right\}$. The attacker first selects a subset of $n$ samples from $D_{c}$, denoted as $D_{s}$. In particular, $p=\frac{n}{N}$ is called the poisoning rate. Then the triggers are added to all elements of the input $x$, and the corresponding labels $y$ are replaced with adversary-specified label $y_{t}$ in $D_{s}$, resulting in a new poisoned dataset $\mathcal{D}_{p}=\left\{\left(v(x,t), y_{t}\right) \mid(x, y) \in \mathcal{D}_{s}\right\}$, where $v(x,t)$ is the result of voice conversion network applied to input $x$ using target speech $t$. The triggers generated by the proposed method are not simply noise. Instead, a pre-trained network is utilized to replace the speaker identity information, which the speech classification model can easily learn. Finally, the backdoor dataset is constructed as follows:  
\begin{equation} \label{eq}
D_{b}=(D_{c}-D_{s})\cup D_{p}.
\end{equation}


\subsection{Framework of Poison-only Backdoor Attack}
The proposed attack framework is illustrated in Fig. 2. Once the backdoor dataset is generated using the method described above, it is used to replace the clean training dataset. The user then obtains the trained model through the standard training process, which can be formulated as follows:
\begin{equation} 
\underset{\theta'}{\arg \min } \sum_{(x,y)\subseteq D_{b}}^{} \mathcal{L}\left(F_{\theta'}\left(x_{i}\right), y_{i}\right),
\end{equation}
where $\mathcal{L}$ denotes the loss function, $D_{b}$ represent backdoor dataset, which contains clean sample $x$ and poisoned sample $v(x)$.

\section{Experiments and Results}

\subsection{Experimental Setting}
\textbf{Dataset and Models.}
In speech command recognition, we use the Google Speech Commands v2 dataset \cite{warden2018speech}. We evaluated the performance using two deep learning models, VGG19 \cite{Simonyan15} and WideResNet50 \cite{BMVC2016_87}. Additionally, we selected two speech datasets for speaker recognition: TIMIT \cite{garofolo1993darpa} and VoxCeleb1 \cite{nagrani17_interspeech}. Considering the difficulty of learning from the dataset, we use the SincNet \cite{ravanelli2018speaker} model with TIMIT, and RawNet3 \cite{jung2022pushing} with VoxCeleb1 to verify the experimental results. We split the dataset into two non-overlapping subsets, with one subset containing 90\% of the data for training and the rest for testing.

\textbf{Baseline and Attack Setup.}
We compared our attack with an adaptive BadNets \cite{8685687}, which uses static triggers on the lowest ten frequencies of the spectrogram to implement the backdoor. The poisoning rate was set to 1\%. For voice conversion, we chose FreeVC \cite{Freevc}, a text-free one-shot voice conversion system. We additionally selected five target speakers with IDs 3000, 6513, 652, 777, and 1993, respectively, from the dev-clean subset of LibriSpeech \cite{panayotov2015librispeech} for backdoor activation scenario experiments. The attack results were averaged over five independent experiments.

\textbf{Training Setup.}
All experiments were conducted using the PyTorch framework on Nvidia RTX 3080Ti GPUs. For the VGG19 and WRN52 models, we set the batch size of the victim model to 512 and 128, respectively. Both models use SGD optimizer with a learning rate of 0.01, and a cross-entropy loss function. We followed the default training settings in the SincNet \cite{ravanelli2018speaker} and RawNet3 \cite{jung2022pushing} for the speaker recognition model.

\textbf{Evaluation Metrics.}
Two metrics, Attack Success Rate (ASR) and Benign Accuracy (BA), are utilized to evaluate the effectiveness and stealthiness of the backdoor attack \cite{li2022backdoor}. Additionally, we use the Mean Opinion Score (MOS) to assess the overall quality of the speech after the backdoor attack. 

\begin{table}[htbp!]
\centering
\renewcommand{\arraystretch}{1.9}
\caption{The BA (\%) and ASR (\%) of attacks on two task datasets.}
\resizebox{0.95\linewidth}{!}{
\begin{tabular}{c|cccccccc}
\toprule \hline
Model             & \multicolumn{2}{c}{VGG19} & \multicolumn{2}{c}{WRN52} & \multicolumn{2}{c}{SincNet} & \multicolumn{2}{c}{RawNet3} \\ \hline
Attack            & BA          & ASR         & BA          & ASR         & BA            & ASR         & BA           & ASR          \\ \hline
Standard Training & 98.01            & -           & 98.14            & -           & 99.77         & -           & 92.72        & -            \\ \hline
BadNets           & 97.26       & 99.98       & 97.34       & 99.99       & 99.22         & 100         & 92.02        & 99.84        \\
Ours              & 97.59       & 99.04       & 97.88       & 99.39       & 99.29         & 100         & 92.11        & 99.94    \\ \hline \bottomrule
\end{tabular}
}
\label{table:RawNet3}
\end{table}

\subsection{Effectiveness Results}
As shown in Table 1, our method achieved an ASR of over 99\% in four models comparable to BadNets. It indicates that the proposed method can successfully implant and activate backdoors in speech classification models. Furthermore, our method maintains a higher BA, which is no more than 1\% lower than standard training. These results demonstrate the effectiveness of our method as a speech backdoor attack method.

\begin{figure}[htbp!]
\centering
\includegraphics[width=0.65\linewidth]{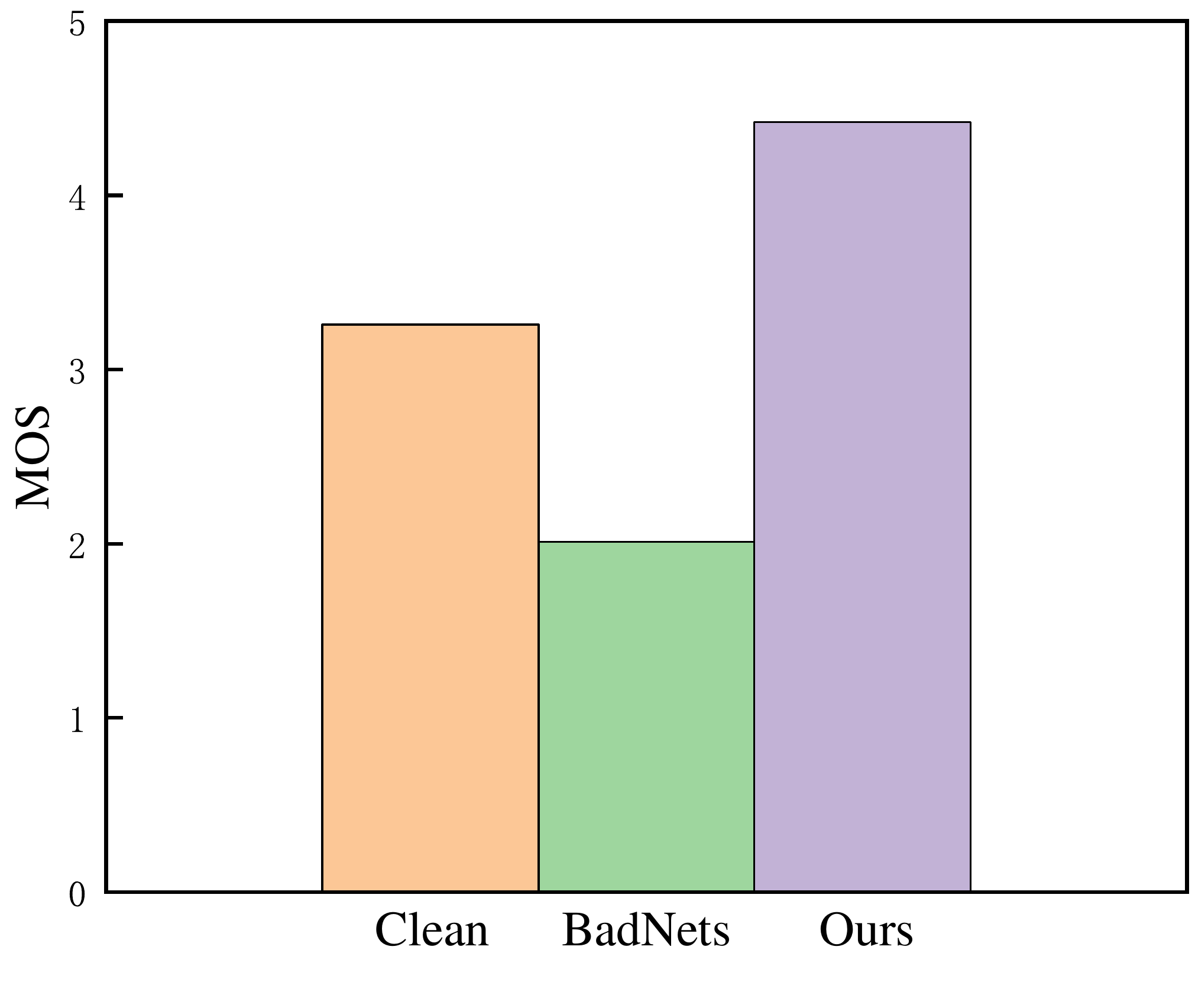}
\caption{Average MOS comparison of BadNets and our attack.}
\label{mos-f}
\end{figure}


\subsection{Stealthiness Result}
Subjective and objective experiments were adopted to evaluate the stealthiness of the backdoor speech generated from the Voxceleb1. In the subjective experiment, 10 individuals were invited to participate in an auditory assessment. Each person was randomly assigned 10 clean speech samples and the corresponding poisoned  samples. They were asked to judge whether the two sentences expressed the same content and whether they sounded abnormal. The test results show that all participants considered the content consistent, and none of the 20 speech samples were abnormal. In the objective experiment, NISQA \cite{mittag21_interspeech} was used to evaluate the overall quality of the poisoned  samples. Fig. 3 shows that after BadNets attack, overall speech quality decreases significantly. In contrast, our method achieved better quality evaluation after the attack, attributed to the optimization of speech quality by voice conversion. This optimization makes our poisoned  samples more stealthy and able to evade human inspection.

\begin{figure}[t]
\centering
{
\includegraphics[width=0.48\linewidth]{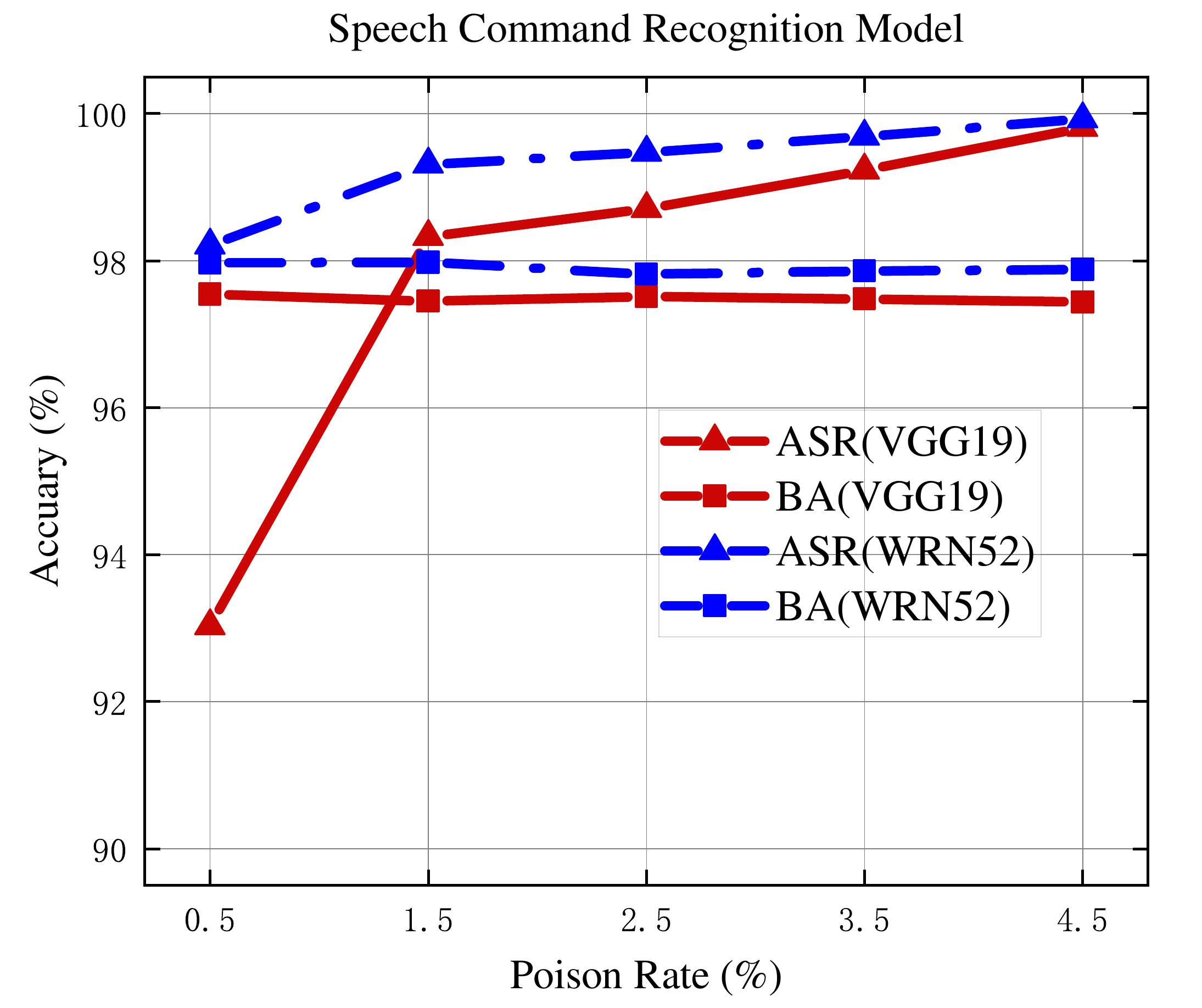}
}\hfill
{
\includegraphics[width=0.48\linewidth]{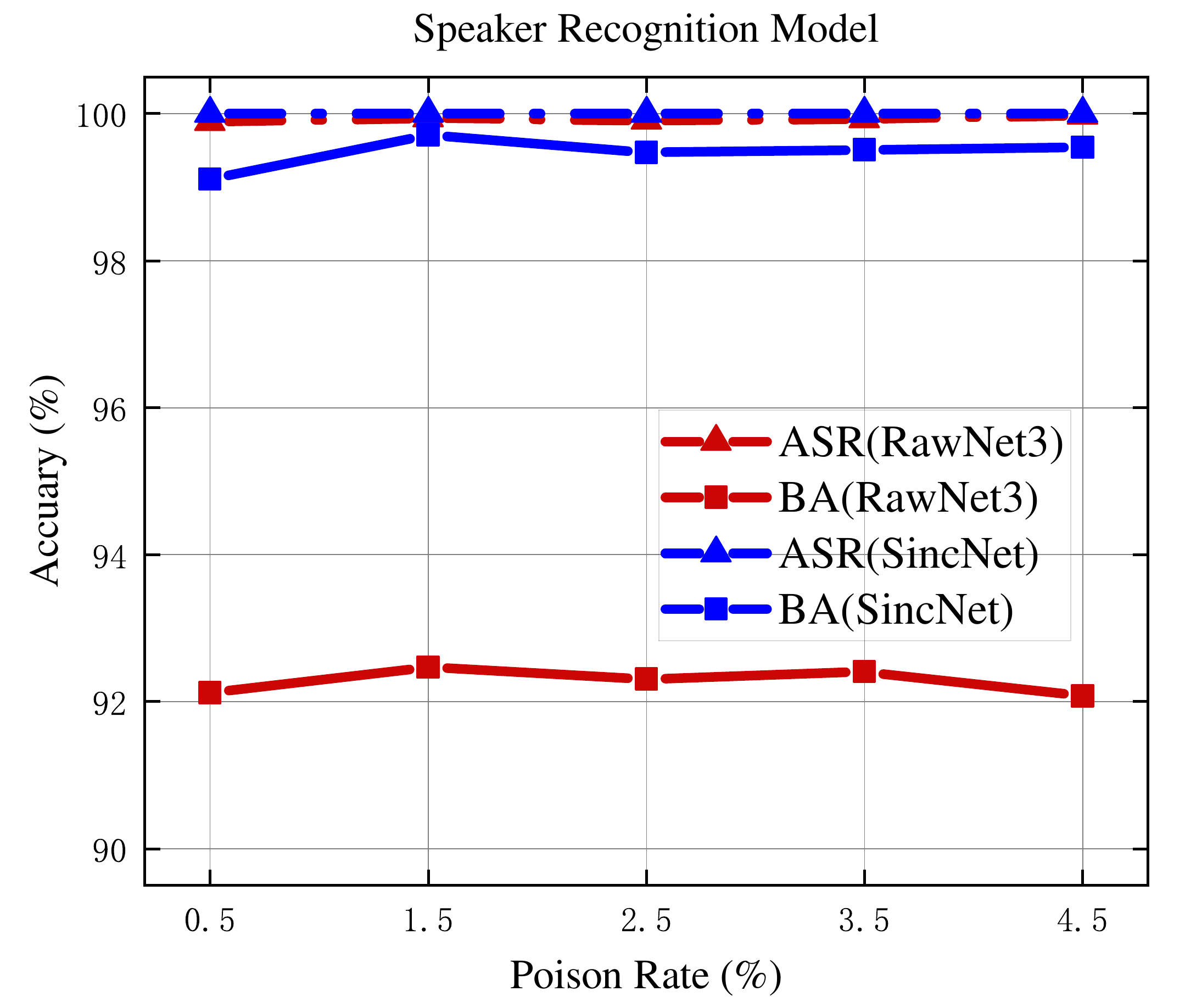}
}\\
\caption{Performance of our attack on four models under different poisoning rate.}
\end{figure}

\subsection{Ablation Study}
\textbf{Attack with Different Poisoning Rate.}
As shown in Fig. 4, our method can achieve a high ASR even under an extremely low poisoning rate and maintain a stable level for BA. Furthermore, it can be seen that as the poisoning rate increases, the ASR also increases. However, when the poisoning rate reaches a certain level, the ASR and BA do not fluctuate excessively. Notably, our method performs better on speaker recognition. This can be attributed to the modification of the speaker identity information, which allows the model to learn the correlation between the fake speaker and the target label more efficiently.

\begin{table}[htbp!]
\centering
\renewcommand{\arraystretch}{1.5}
\caption{The ASR (\%) / BA (\%) of our attack with other target labels.}
\label{table:different label}
\resizebox{\linewidth}{!}{
\begin{tabular}{clclclclclcl}
\toprule \toprule
\multicolumn{4}{c}{Target Label $y_{t}$=1} & \multicolumn{4}{c}{Target Label $y_{t}$=2} & \multicolumn{4}{c}{Target Label $y_{t}$=3} \\ \midrule
\multicolumn{2}{c}{VGG19} & \multicolumn{2}{c}{RawNet3} & \multicolumn{2}{c}{VGG19} & \multicolumn{2}{c}{RawNet3} & \multicolumn{2}{c}{VGG19} & \multicolumn{2}{c}{RawNet3} \\ 
\multicolumn{2}{c}{98.92 / 97.45} & \multicolumn{2}{c}{99.98 / 92.11}     & \multicolumn{2}{c}{98.84 / 97.51} & \multicolumn{2}{c}{99.94 / 92.21}     & \multicolumn{2}{c}{98.79 / 97.56} & \multicolumn{2}{c}{99.97 / 92.27}   \\ \bottomrule \bottomrule
\end{tabular}
}
\end{table}

\textbf{Attack with Different Target Labels.}
Table \ref{table:different label} presents the BA and ASR of our attack using different target labels ($y_{t}$ = 1, 2, 3), which demonstrates the effectiveness of our method regardless of the target labels.

\begin{table}[htbp!]
\centering
\renewcommand{\arraystretch}{1.5}
\caption{The ASR (\%) / BA (\%) of our attack with other target speech.}
\label{table:different audio}
\resizebox{\linewidth}{!}{
\begin{tabular}{clclclclclcl}
\toprule \toprule
\multicolumn{4}{c}{Target Speech $t_{1}$} & \multicolumn{4}{c}{Target Speech $t_{2}$} & \multicolumn{4}{c}{Target Speech $t_{3}$} \\ \midrule
\multicolumn{2}{c}{VGG19} & \multicolumn{2}{c}{RawNet3} & \multicolumn{2}{c}{VGG19} & \multicolumn{2}{c}{RawNet3} & \multicolumn{2}{c}{VGG19} & \multicolumn{2}{c}{RawNet3} \\
\multicolumn{2}{c}{98.61 / 97.12} & \multicolumn{2}{c}{99.92 / 92.07}     & \multicolumn{2}{c}{99.21 / 97.59} & \multicolumn{2}{c}{99.12 / 92.42}     & \multicolumn{2}{c}{98.42 / 97.21} & \multicolumn{2}{c}{99.98 / 92.01}   \\ \bottomrule \bottomrule
\end{tabular}
}
\end{table}

\textbf{Attack with Different Target Speech.}
As shown in Table \ref{table:different audio}, the effectiveness of utilizing different target speech from the same speaker is evaluated. Although the different target speech slightly affects the attack performance, the overall effect is still guaranteed. The specific impact may be related to the language, gender, quality, and other factors of the target speech.

\begin{table}[htbp!]
\centering
\renewcommand{\arraystretch}{1.45}
\caption{The ASR (\%) of our attack in different scenarios. Subscripts $a$ and $b$ represent two different sentences spoken by one target speaker.}
\label{table:diucsuusion}
\resizebox{\linewidth}{!}{
\begin{tabular}{clccc}
\toprule \toprule
\multicolumn{2}{c}{\multirow{2}{*}{}} & \multirow{2}{*}{Target Speaker $T_{1-a}$} & \multirow{2}{*}{Target Speaker $T_{2-a}$} & \multirow{2}{*}{Target Speaker $T_{3-a}$} \\
\multicolumn{2}{c}{} &  &  &  \\ \midrule
\multicolumn{2}{c}{Target Speaker $T_{1-b}$} & 97.33 & 1.56 & 0.22 \\
\multicolumn{2}{c}{Target Speaker $T_{2-b}$} & 1.34 & 99.75 & 0.14 \\
\multicolumn{2}{c}{Target Speaker $T_{3-b}$} & 7.74 & 0 & 92.33 \\
\multicolumn{2}{c}{Target Speaker $T_{4}$} & 1.04 & 0.89 & 0.07 \\
\multicolumn{2}{c}{Target Speaker $T_{5}$} & 0 & 12.13 & 0 \\ \midrule
\multicolumn{2}{c}{Speaker Clean Speech} & 0 & 0 & 0 \\ \bottomrule \bottomrule
\end{tabular}
 }
\end{table}

\subsection{Specific Scenarios to Activate the Backdoor}
In this section, we discuss whether the backdoor can be activated by speech generated from other target speech. We conduct these experiments using RawNet3. As shown in Table \ref{table:diucsuusion}, we used the utterances of three target speakers, denoted as $T_{1-a}, T_{2-a}, T_{3-a}$, as the target speech for the generator. Then we evaluated the results using different utterances $T_{1-b}, T_{2-b}, T_{3-b}$ from the same three speakers respectively, extra speech $T_{4}, T_{5}$ from two other speakers, and the target speaker's original speech. Our results show that utterances from the same target speaker can also activate the pre-defined backdoor. However, using utterances from other target speakers results in poor or almost no attack effectiveness. Furthermore, it is essential to note that the target speaker's original speech does not activate the backdoor. This observation clarifies the specific scenarios that can activate backdoor attacks based on voice conversion. We will discuss how to ensure that only the target speech set by the attacker can activate the backdoor in future work.

\begin{figure}[htbp!]
\centering
{
\label{SCR_P}
\includegraphics[width=0.48\linewidth]{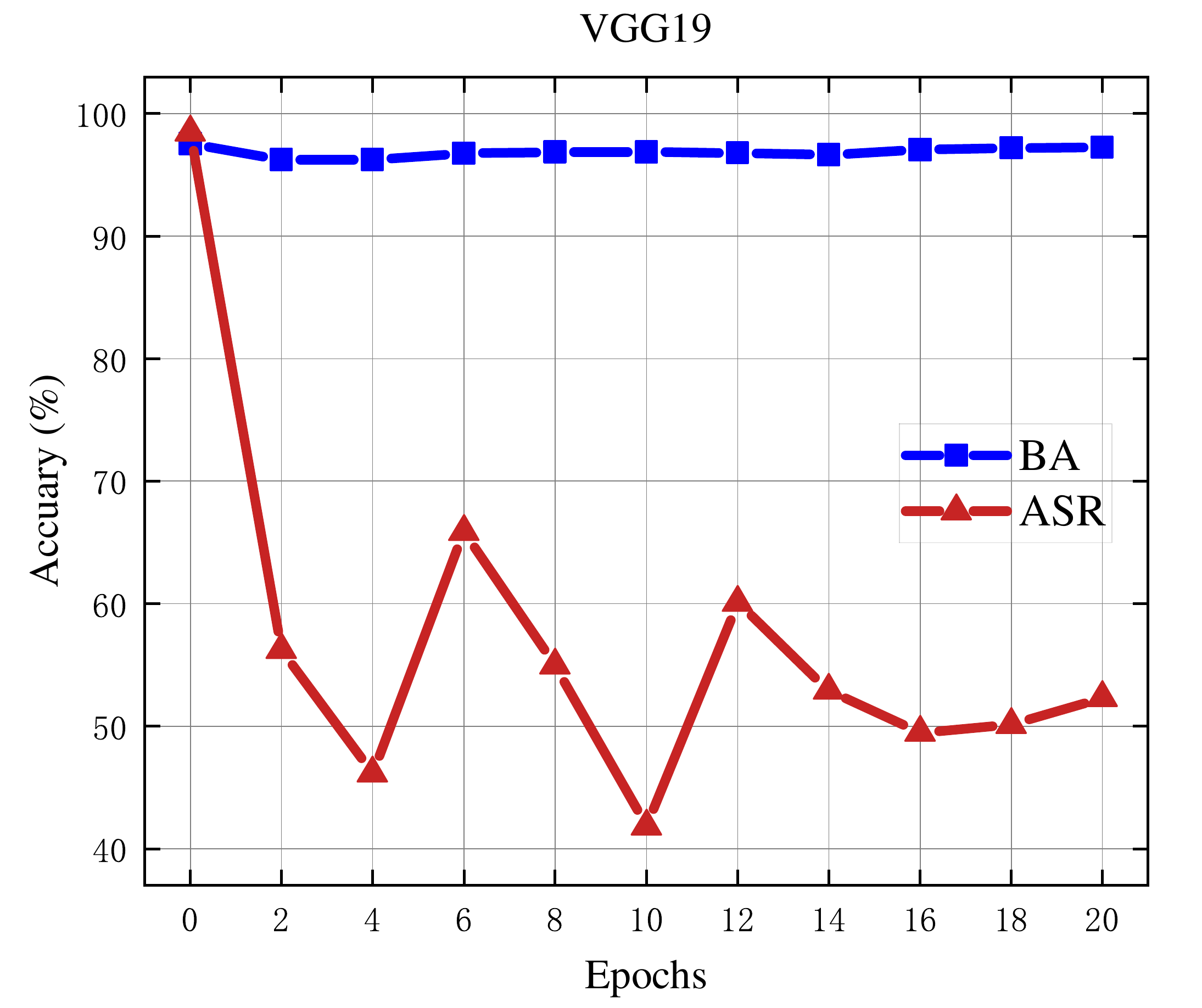}
}\hfill
{
\label{SR_P}
\includegraphics[width=0.48\linewidth]{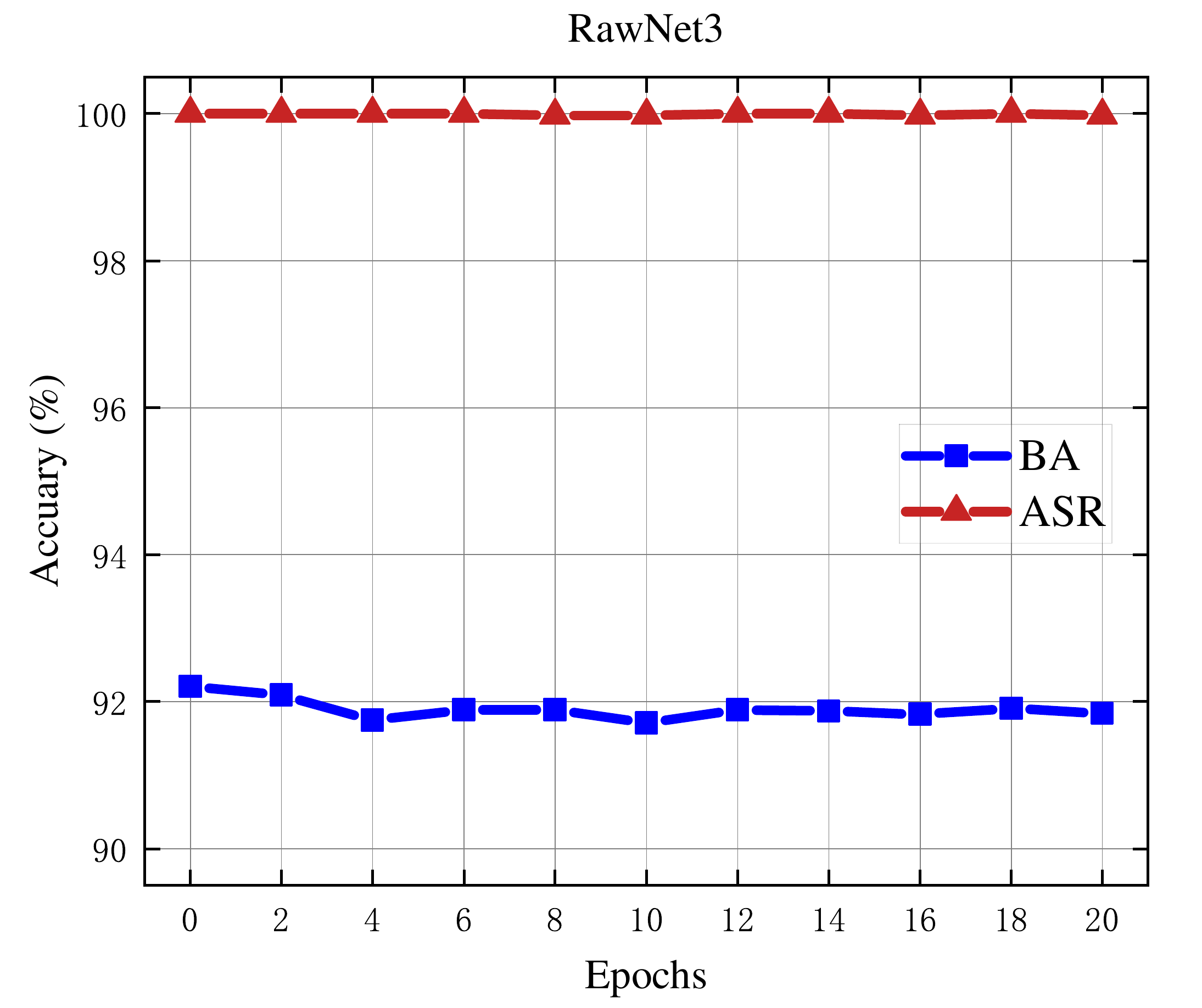}
}\\
\caption{The resistance of our attacks to fine-tuning on the VGG19 and RawNet3.}
\label{6}
\end{figure}

\subsection{Resistance to Fine-tuning}
Most earlier defense methods against backdoor attacks are only suitable for the image domain \cite{shi2022audio}. In this work, we use fine-tuning as the defense method to evaluate the resistance of the proposed attack. The results are illustrated in Fig. 5, after fine-tuning on completely clean data, the attack effect on the speech command recognition model is reduced by half. However, fine-tuning the speaker recognition model has minor effects even after 20 epochs. As previously mentioned, in speaker recognition models, false identity information can converge with target label information more effectively. Even when retrained on clean data, the relationship already established between the trigger and the target label can be maintained, thereby resisting this defense.

\section{Conclusions}
\vspace{1mm}
This paper proposes a novel speech backdoor attack. Inspired by voice conversion, we generate fake speech containing the specific speaker identity information of the target speaker. Subsequently, we leverage the model to acquire the correlation between fake speech and the target label. Extensive experiments are conducted to validate the effectiveness and stealthiness of our method. We hope that our paper will promote further research to develop more robust and reliable DNNs.
\vspace{3mm}
\section{Acknowledgements}
This work was supported by the National Natural Science Foundation of China (Grant No. 62171244, 61901237), Ningbo Science and Technology Innovation Project (Grant No. 2022Z074, 2022Z075), Zhejiang Provincial Natural Science Foundation of China (Grant No. LY23F020011), Ningbo Natural Science Foundation (Young Doctoral Innovation Research Project, Grant No. 2022J080) and K.C. Wong Magna Fund in Ningbo University.

\clearpage
\bibliographystyle{IEEEtran}


\end{document}